\documentclass [12pt] {article}
\usepackage{amsmath}   
\usepackage{graphicx}   
\usepackage{verbatim}   
\usepackage{color}      
\usepackage{subfigure} 
\usepackage{hyperref}
\usepackage{latexsym}
\usepackage{caption}
\usepackage{wrapfig}
\usepackage{geometry}
\begin{document}
\begin{center}
\textbf{Fundamental Constants as Monitors of the Universe}\\
\end{center}

\begin{center}
Rodger I. Thompson\\
\end{center}

\noindent Department of Astronomy and Steward Observatory, University of Arizona,\\
Tucson, Arizona 85721, USA\\
E-mail: rit@email.arizona.edu

\begin{abstract}
Astronomical observations have a unique ability to determine the laws of physics at distant times
in the universe.  They, therefore, have particular relevance in answering the basic question as to
whether the laws of physics are invariant with time.  The dimesionless fundamental constants, such as
the proton to electron mass ratio and the fine structure constant are key elements in the investigation.
If they vary with time then the answer is clearly that the laws of physics are not invariant with time and
significant new physics must be developed to describe the universe.  Limits on their variance, on the
other hand, constrains the parameter space available to new physics that requires a variation with
time of basic physical law. There are now observational constraints on the time variation of the proton 
to electron mass ratio $\mu$ at the $10^{-7}$ level.  Constraints on the variation of the fine structure 
constant $\alpha$ are less rigorous $(10^{-5})$ but are imposed at higher redshift.  The implications
of these limits on new cosmologies that require rolling scalar fields has already had its first investigations.
Here we address the implications on basic particle physics.  The proton to electron mass ratio is obviously
dependent on the particle physics parameters that set the mass of the proton and the electron.  To first
order the ratio is dependent on a combination of the Quantum Chromodynamic scale,  the Yukawa couplings, 
and the Higgs Vacuum Expectation Value.  Here that relationship is quantitative defined for the first time.
When coupled with previous determinations of the relation of the fine structure constant to the same 
parameters two constraints exist on the fractional variation of these parameters with time.  A third
independent constraint involving only the three parameters could set the stage for constraints on their
individual fractional variation.
\end{abstract}

keywords: Cosmology, Fundamental Constants, Particle Physics

\section{Introduction}\label{s-intro}
\setcounter{figure}{0}
The considerable evidence for late time acceleration of the expansion of the universe has
also accelerated considerations of cosmologies other than the conventional $\Lambda$CDM
cosmology and new physics beyond the standard model.  Inherent in many of these alternative
cosmologies and new physics is the possibility that fundamental constants such as the proton
to electron mass ratio $\mu$ and the fine structure constant $\alpha$ may roll with time. The
cosmological implications of such variations in time has previously been discussed in~\cite{cal11}, 
~\cite{thm12} and ~\cite{thm13}, therefore the emphasis here is on the implications for new physics.
Variations of $\mu$ and $\alpha$ requires variations in basic physics parameters 
such as the Quantum Chromodynamic scale, $\Lambda_{QCD}$,  the Yukawa couplings, $h$ and 
the Higgs Vacuum Expectation Value $\nu$.  The expected time scales for a variation of the constants
is on the order of a Hubble time which points to astronomical observations as the natural mode
for detecting any change.  The implication of a change in $\alpha$  for the parameters has been
previously explored in Ref.~\cite{coc07} and \cite{luo11}.  Building on previous work such as e.g.  
Refs.~\cite{cam95}, \cite{lan02}, \cite{dam03}, \cite{coc07}, \cite{luo11}, \cite{fer12} and  
\cite{fer14} a connection between a variance of $\mu$ and the parameters is developed.   A limit 
on the variation of $\mu$ at a redshift of 0.89 establishes the stability of $\mu$ to a part in $10^7$ 
over a time period of a bit more than the last half of the age of the universe.  Combined with the limits 
on (or observation of) a change in $\alpha$ these observations define the allowed parameter space for 
deviations from the Standard Model of physics, a subject with few observational constraints. In 
section~\ref{s-maobs} the current observational limits on the fractional variations of $\mu$ and 
$\alpha$ are discussed followed by the development of the relation betwen the fractional variation
of $\mu$ and the fractional variations of $\Lambda_{QCD}$, $\nu$ and $h$.  The previously determined
relation of $\alpha$ is introduced to produce a set of two independent equations for the variation of th
three parameters and the possiblity of a third independent relationship is discussed.

\section{$\mu$ and $\alpha$ Observations} \label{s-maobs}
Astronomical spectroscopic observations that set limits on the variation of $\mu$ and $\alpha$
are generally observations of cold gas in Damped Lyman Alpha clouds (DLAs) that lie along the
line of sight to a quasar.  This gas produces absorption lines that are generally narrow and
whose wavelengths can be established with reasonable accuracy.  As its name implies, $\alpha$
measurements are made by observing atomic fine structure transitions in several different elements.
Measurements of $\mu$, on the other hand, involve searching for shifts in the spectra of molecular 
species such as molecular hydrogen Ref.~\cite{thm75}.  While the $\alpha$ observations are all
at optical wavelengths the $\mu$ measurements include both optical and radio observations.
The higher spectral resolution of radio measurements provides very stringent constraints on a
variation of $\mu$.  The situation with regard to $\alpha$ is a bit murky.  There are reports of
both temporal and spatial changes in $\alpha$ (Ref.~\cite{web01}, Ref.~\cite{web11}) at the $10^{-5}$
level which have been a subject of discussion.  In contrast recent dedicated VLT with UVES 
observations (Ref.~\cite{eva14}, Ref.~\cite{mol13}) have not been able to confirm these reports.

\subsection{$\mu$ measurements} \label{ss-muobs}
\begin{figure}
  \vspace{30pt}
\scalebox{.5}{\includegraphics{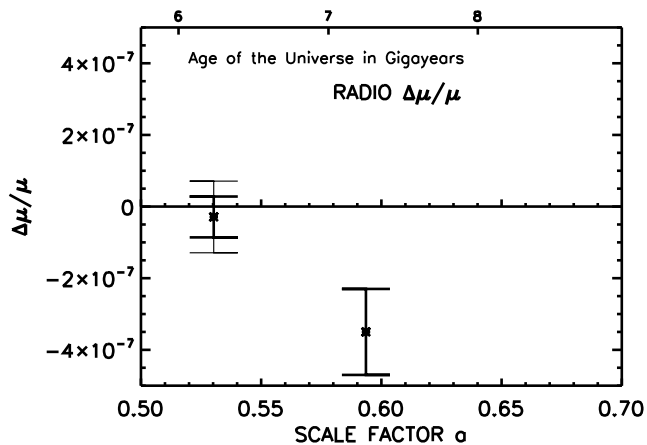}}
  \caption{The low redshift radio $\Delta \mu / \mu$ constraints at z = 0.6874 and z = 0.88582 plotted
versus the scale factor $a = 1/(1+z)$.  The error bars are $1\sigma$, however, an additional larger error bar 
representing systematic effects is over plotted on the z = 0.88582 ($a=0.53$) constraint.  These are
 the primary constraints utilized in this letter. The age of the universe in gigayears is plotted on the top
axis and in Fig.~\ref{fig-allerr}.} 
\label{fig-raderr}
\end{figure}

\begin{figure}
  \vspace{30pt}
\scalebox{.5}{\includegraphics{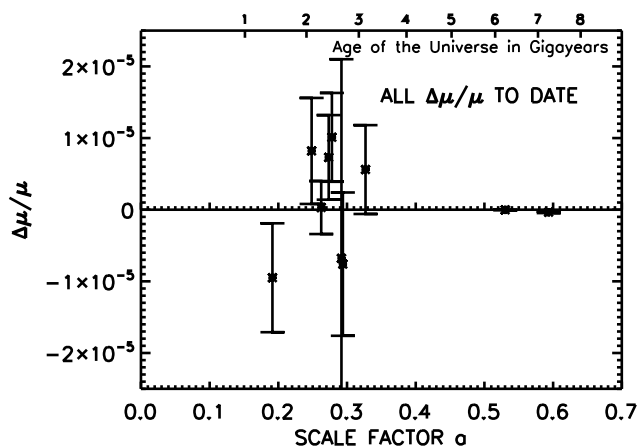}}
  \caption{All of the observational constraints on $\Delta \mu / \mu$ from radio ($z < 1$)
and optical ($z > 1$) observations  plotted versus the scale factor $a = 1/(1+z)$.  All constraints are 
at the $1 \sigma$ level. The low redshift radio constraints are difficult to see at the scale of this plot.} 
\label{fig-allerr}
\end{figure}

Observational constraints on $\frac{\Delta \mu}{\mu}$ exist over the redshift range from 0.68 to
4.2.  The most stringent limits come from radio observations of methanol and ammonia absorption
lines in two systems with redshifts less than one (Ref.~\cite{kan11}, Ref.~\cite{kan15}).   These are the limits used here 
but for completeness Table~\ref{tab-ob} lists the most stringent $\frac{\Delta \mu}{\mu}$ constraints 
for the ten objects examined for changes in $\mu$.  Fig.~\ref{fig-raderr} shows the two low redshift
observations in detail and Fig.~\ref{fig-allerr} shows all of the observations.  The radio observations
errors are barely visible in Fig.~\ref{fig-allerr} indicating the significantly improved constraints that the radio 
observations provide.
\begin{table}
\begin{tabular}{llllll}
\hline
\hline
Object & Redshift  & $\Delta \mu / \mu$ & $1\sigma$ error  & Ref.\\
\hline
J1443+2724 & 4.224 & $ -9.5 \times 10^{-6}$ & $\pm 7.6 \times 10^{-6}$ & Ref.~\cite{bag15}\\
Q0347-383 & $3.0249$ & $8.2 \times 10^{-6}$ & $\pm 7.4\times 10^{-6}$  & Ref.~\cite{kin08}\\
Q0528-250 & $2.811$ & $3.0 \times 10^{-7}$ & $\pm 3.7 \times 10^{-6}$  &  Ref.~\cite{kin11}\\ 
J1237+0647 & $2.689$ & $-5.4 \times 10^{-6}$ & $\pm 7.5 \times 10^{-6}$ & Ref.~\cite{dap15}\\
Q J0643-5041 & $2.659$ & $7.3 \times 10^{-6}$ & $\pm 5.9 \times 10^{-6}$ & Ref.~\cite{rah14}\\
Q0405-443 & $2.5974$ & $10.1 \times 10^{-6}$ & $\pm 6.2 \times 10^{-6}$ &  Ref.~\cite{kin08}\\
Q2348-011 & $2.426$ & $-6.8 \times 10^{-6}$ & $\pm 27.8 \times 10^{-6}$ & Ref.~\cite{bag12}\\
He0027-1836 & $2.402$ & $-7.6 \times 10^{-6}$ & $\pm 1.0 \times 10^{-5}$ &  Ref.~\cite{rah13}\\
J2123-005 & $2.059$ & $5.6 \times 10^{-6}$ & $\pm 6.2 \times 10^{-6}$ &  Ref.~\cite{mal10}\\
PKS1830-211 & $0.88582$ & $-2.9 \times 10^{-8}$ & $\pm 5.7 \times 10^{-8}$ &  Ref.~\cite{kan15}\\
B0218+357 & $0.6847$ & $-3.5 \times 10^{-7}$ & $\pm 1.2 \times 10^{-7}$ &  Ref.~\cite{kan11}\\
\hline
\hline
\end{tabular}
\caption{Current best observational constraints on $\frac{\Delta \mu}{\mu}$}   \label{tab-ob}
\end{table}

To date there is no substantiated $(>5\sigma)$ $\mu$ variation at any
redshift.  Changes in $\mu$ are detected by changes in molecular spectra Ref.~\cite{thm75} which 
accounts for the paucity of objects since molecular absorption features are found in only a few 
extragalactic sources.  The PKS1830-211 methanol radio observation Ref.~\cite{kan15}
is the most stringent constraint on a change in $\mu$ or at $\frac{\Delta \mu}{\mu}
= (-2.9 \pm 5.7) \times 10^{-8}$.  This is the statistical limit using all four of the observed methanol
absorption lines which includes the 12.179 GHz line.  Doubts are expressed in Ref.~\cite{kan15} that
this line has a completely common line of sight with the other three lines increasing the 
uncertainty to $10^{-7}$.  The statistical result is retained in Table~\ref{tab-ob} and Fig.~\ref{fig-allerr}
 but a $1\sigma$ uncertainty of $\frac{\Delta \mu}{\mu} = 10^{-7}$ is utilized in this
analysis and Fig.~\ref{fig-raderr}. This new constraint is a primary motivation for exploring the
constraints on other fundamental physical parameters.  

The new limits on $\frac{\Delta \mu}{\mu}$ are at redshifts less than one, however, this time period 
is 57\% of the age of a flat universe with $H_0=70$ and $\Omega_m=0.3$.  For comparison the bound on  
$\frac{\Delta \mu}{\mu}$ at $z=0.88582$ is equivalent to a linear time evolution of less than $\pm 7.88 
\times 10^{-18}$ per year as compared to current atomic  clock measurements of $\pm 5.8 \times 
10^{-16}$ per year Ref.~\cite{fer12}.   A linear evolution, however is very unlikely and is just used here 
as a comparison to the laboratory value.

\subsection{$\alpha$ measurements} \label{ss-aobs}
It is beyond the scope of this article to thoroughly discuss the current state of $\alpha$ measurements.
If the claims of a temporal and spatial variation of $\alpha$ by Ref.~\cite{web11} are true it would require a
major deviation from the standard model.  For the purposes of this paper $\alpha$ is assumed to be 
stable at approximately the $10^{-5}$ level consistent with Ref.~\cite{mol13} and Ref.~\cite{eva14}.  This 
does not affect the main result of the paper which is the connection between $\mu$ and the parameters.

\section{The dependence of $\mu$ on $\Lambda_{QCD}$, $\nu$ and h$_e$} \label{s-hqy}
The new stringent constraints on $\frac{\Delta \mu}{\mu}$ provide a strong motivation to
determine the equivalent constraints they impose on the fundamental physics parameters that
set  the proton and electron masses.  In the standard model these parameters do not vary
but in super symmetry (SUSY) and other new physics theories they can.  The constraints on
the variance of $\mu$ therefore determine how far the new theories can stray from the standard 
model.  The variation of $\mu$ is simply
\begin{equation}\label{eq-dmumu}
\frac{d\mu}{\mu}=\frac{dm_p}{m_p}-\frac{dm_e}{m_e}
\end{equation} 
The proton mass variation involves all three of the fundamental parameters Ref.~\cite{coc07}.
\begin{equation} \label{eq-mpf}
\frac{dm_p}{m_p} = a \frac{d\Lambda_{QCD}}{\Lambda_{QCD}} + b (\frac{dh_e}{h_e}+\frac{d\nu}{\nu})
\end{equation}
where $a$ and $b$ are constants of order unity and $h_e$ is the electron Yukawa coupling.  In eqn.~\ref{eq-mpf} 
the coefficients of the right hand terms have been left to be the general coefficients $a$ and $b$ but 
in Ref.~\cite{coc07} they have been set to add to one, $a=0.76$ and $b=0.24$ .  This is because the proton mass 
is assumed to be proportional to a factor
given by $(\Lambda_{QCD})^a(h_e \nu)^b$.  Both $\Lambda_{QCD}$ and $\nu$ have the units of
mass while the Yukawa coupling $h_e$ is dimensionless.  Since the product must have the dimensions
of mass to match the proton mass the powers $a$ and $b$ must add to one.  When the logarithmic
derivatives are taken on both sides the powers become the coefficients and therefore must add to
one as they do in Ref.~\cite{coc07}.  If, however, there was another term in the proportionality that had the units of
mass but did not vary then it would not appear in eqn.~\ref{eq-mpf} and the coefficients would not
add to one.  For this reason the coefficients  are left as the general $a$ and $b$. 

By definition the electron mass is $\nu h_e$, the product of the Higgs VEV and the electron Yukawa 
coupling, therefore
\begin{equation} \label{eq-dme}
\frac{dm_e}{m_e}=\frac{dh_e}{h_e} + \frac{d\nu}{\nu}
\end{equation} 
Combining eqns.~\ref{eq-dmumu},  ~\ref{eq-mpf} and ~\ref{eq-dme}  yields
\begin{equation} \label{eq-qhy}
\frac{d\mu}{\mu} = a\frac{d\Lambda_{QCD}}{\Lambda_{QCD}}+ (b-1)( \frac{dh}{h}+\frac{d\nu}{\nu})
\end{equation}
using the common assumption that the logarithmic derivatives $\frac{dh_i}{h_i}$ are  the  same for all 
$i$ (Ref.~\cite{coc07}).  If $a$ plus $b$ add to one then the much cleaner expression in eqn.~\ref{eq-one} is true.
\begin{equation} \label{eq-one}
\frac{d\mu}{\mu} = a(\frac{d\Lambda_{QCD}}{\Lambda_{QCD}}- (\frac{dh}{h}+\frac{d\nu}{\nu}))
\end{equation}

Eqn.~\ref{eq-qhy} provides the connection between a change in the dimensionless fundamental constant
$\mu$ and a combination of changes in $\Lambda_{QCD}$, $h$ and $\nu$.  It assumes that the standard
model relation between the parameters and the masses of the proton and electron still hold but unlike
the standard model they are allowed to vary.  The connection is, of course, consistent with the  standard
model where all of the terms are zero.  Note that this is different from recent work by Ref.~\cite{cal14} 
and Ref.~\cite{fri15} who just considered a variation of the proton mass and restricted its change to only a
variation of $\Lambda_{QCD}$.  The observational limit on $\Delta \mu /\mu$ then gives
\begin{equation} \label{eq-mucon}
a\frac{d\Lambda_{QCD}}{\Lambda_{QCD}} +(b-1)( \frac{dh}{h}+\frac{d\nu}{\nu}) ~< 10^{-7}
\end{equation}
over the last $57\%$ of the age of the universe.  The previously derived connection for $\alpha$ (Ref.~\cite{coc07},
Ref.~\cite{luo11}) is
\begin{equation} \label{eq-alpha}
R \frac{\Delta \alpha}{\alpha} = \frac{\Delta \Lambda_{QCD}}{\Lambda_{QCD}}-\frac{2}{27}(3\frac{\Delta \nu}{\nu}+\frac{\Delta h_c}{h_c}+\frac{\Delta h_b}{h_b}+\frac{\Delta h_t}{h_t})
\end{equation}
where R is a model dependent scalar.  Under the assumption that $\frac{\Delta h_i}{h_i} = \frac{\Delta h}{h}$
it becomes
\begin{equation} \label{eq-ah}
 \frac{\Delta \alpha}{\alpha} = \frac{1}{R}[\frac{\Delta \Lambda_{QCD}}{\Lambda_{QCD}}-\frac{2}{9}(\frac{\Delta \nu}{\nu}+\frac{\Delta h}{h})] ~< 10^{-5} 
\end{equation}
Equations~\ref{eq-mucon} and~\ref{eq-ah} relate fractional changes in $\mu$ and $\alpha$ to fractional 
changes in $\Lambda_{QCD}$, $\nu$ and $h$.  The far right hand term in each of the relations represents 
the current observational constraints on $\mu$ and $\alpha$.

Equations~\ref{eq-mucon} and~\ref{eq-ah} are two equations in the three unknowns 
$\Delta \Lambda_{QCD}/\Lambda_{QCD}$, $\Delta \nu/\nu$, and $\Delta h/h$.  As such they allow
the ellimination of one of the variables but the resultant is still the combination of two of the unknowns.
A third independent equation would allow the separation of the unknowns into three separate constraints
on the individual parameters.  A possibility is the gyromagnetic ratio $g$ which can be expressed in a 
Taylor series expansion in $\alpha$.  Hyperfine splitting, such as observed in the 21 cm line is dependent
on $g$ along with $\alpha$ and $\mu$ which could provide an observational channel for the third equation.

\section{Conclusions}
Fundamental constants such as $\mu$ and $\alpha$ are sensitive monitors of possible changes
in time of the basic laws of physics.  Both involve complex physics that have many components
and inter-relationships such that the variance of any of the parameters can not be considered in
isolation from the rest.  The fundamental constants, however, are pure numbers and changes in
those numbers are unambiguous and measurable.  If the standard model relationships between the
fundamental constants and the dimensionfull parameters are maintained fractional changes in the 
constants establish quantitative connections to fractional changes in the parameters.  This work 
provides the connection between fractional changes in $\mu$ and a combination of fractional changes
in the quantum chromodynamic scale, the Higgs vacuum expectation energy and the Yukawa
couplings.  This connection has model dependent parameters as does the previously established
connection with fractional changes in $\alpha$.  A third possible constraint involves the 
gyromagnetic ratio $g$.  Three indepedent equations raises the possibility of constraining the
three parameters separately.  To date no confirmed change with time has been
found for either $\mu$ or $\alpha$ on astronomical time scales.  These observational limits
then provide quantitative limits on the allowed variance of the combination of parameters
but not on the individual parameters themselves.  The observational limits are consistent with the 
standard model and provide limits on the allowed deviations from that model.

\section{Acknowledgements}

The author thanks Jullian Berengut for enlightening discussions on fundamental constant
variation versus parameter variation and  Keith Olive for discussions clarifying the dimensional argument 
of why the coefficients in equation 11 of Ref.~\cite{coc07} should add to one. C. J. A. P. Martins is thanked for
useful discussions in preparing this article.  The author also benefited from discussions  with D. Psaltis,  D. 
Arnett, A. Coc and P. Pinto.

\end{document}